\documentclass[letterpaper,twocolumn,10pt]{article}
\usepackage{usenix2019_v3}
\usepackage[titlenumbered, linesnumbered,vlined, ruled, boxed]{algorithm2e}
\usepackage{algorithmicx}
\usepackage{algpseudocode}
\usepackage{tabularx}
\usepackage{multirow}
\usepackage{graphicx}
\usepackage{capt-of}
\usepackage[super]{nth}
\usepackage{placeins}

\newcommand{\kmeans}{\emph{k}-means}

\newcolumntype{Y}{>{\centering\arraybackslash}X}
\newcolumntype{P}[1]{>{\centering\arraybackslash}p{#1}}

\newcommand{\etal}{\textit{et al}.}
\newcommand{\ie}{\textit{i}.\textit{e}.}
\newcommand{\eg}{\textit{e}.\textit{g}.}

\SetKwRepeat{Do}{do}{while}

\newcommand{\toolname}{SYNAPSE}

\begin{document}

\date{}

\title{\Large \bf Processing Tweets for Cybersecurity Threat Awareness}

\author{
{\rm Fernando Alves, Aurélien Bettini, Pedro M. Ferreira, and Alysson Bessani}\\
LASIGE, Faculdade de Ciências, Universidade de Lisboa -- Portugal}

\maketitle

\begin{abstract}
Receiving timely and relevant security information is crucial for maintaining a high-security level on an IT infrastructure.
This information can be extracted from Open Source Intelligence published daily by users, security organisations, and researchers.
In particular, Twitter has become an information hub for obtaining cutting-edge information about many subjects, including cybersecurity.
This work proposes \toolname{}, a Twitter-based streaming threat monitor that generates a continuously updated summary of the threat landscape related to a monitored infrastructure.
Its tweet-processing pipeline is composed of filtering, feature extraction, binary classification, an innovative clustering strategy, and generation of Indicators of Compromise (IoCs).
A quantitative evaluation considering all tweets from 80 accounts over more than 8 months (over 195.000 tweets), shows that our approach timely and successfully finds the majority of security-related tweets concerning an example IT infrastructure (true positive rate above 90\%), incorrectly selects a small number of tweets as relevant (false positive rate under 10\%), and summarises the results to very few IoCs per day.
A qualitative evaluation of the IoCs generated by \toolname{} demonstrates their relevance (based on the CVSS score and the availability of patches or exploits), and timeliness (based on threat disclosure dates from NVD).
\end{abstract}

\section{Introduction}

A security analyst must be aware of the latest developments regarding updates, patches, mitigation measures, vulnerabilities, attacks, and exploits to adequately protect an IT infrastructure.
\emph{Security Operations Centers} (SOC) improve their awareness through \emph{Security Information and Event Management} (SIEM) software, thereby allowing the correlation of the latest cybersecurity developments with internal infrastructure events.

There are two primary ways of obtaining cybersecurity news.
One is to purchase a curated feed from a specialised company such as SenseCy~\cite{sensecy} or SurfWatch~\cite{surfwatch}.
Another, is to collect \emph{Open Source Intelligence} (OSINT)~\cite{steele1996open} available from various sources on the internet (\eg{}, Threatpost~\cite{threatpost}).

There are numerous \emph{threat intelligence} tools (\eg{}, SpiderFoot~\cite{spiderfoot}, IntelMQ~\cite{intelmq}) that can collect security-related OSINT from a wide variety of sources, including such feeds.
However, these use simple keyword-based filters to narrow the big volume of collected information, not employing any sophisticated methodology to select only the relevant data or handling data aggregation and duplicate removal---two fundamental characteristics for an efficient OSINT usage~\cite{sauerwein2017threat}.
Moreover, recent work (\eg{},~\cite{liao2016acing,zhu2016featuresmith,sabottke2015vulnerability}) demonstrates that different types of useful information and Indicators of Compromise (IoC) can be obtained from OSINT through the application of machine learning techniques.
These results highlight the gap between the current capabilities of existing OSINT-processing tools and the intelligence OSINT can provide.

To address this gap, this paper proposes \toolname{}, a Twitter-based streaming threat monitor that generates a continuously updated summary of the threat landscape concerning a monitored infrastructure.
\toolname{} gathers tweets from carefully selected security-related accounts, selects those relevant for the specified monitored infrastructure using supervised machine learning, and avoids presenting repeated information by employing a novel stream clustering method.

\toolname{}'s design addresses three main challenges: OSINT collection, cybersecurity-related content selection, and the aggregation of related tweets through a stream clustering algorithm adapted to the context of cybersecurity.
A threat intelligence tool must address the first two challenges to ensure its usefulness, \ie{}, it must collect large amounts of data and accurately select those that are relevant for the SOC.
The aggregation challenge is paramount to promote the efficient operation of the SOC, \ie{}, the analysts have a limited time budget to evaluate the current threat landscape, thus the tool must present only a summary of the most relevant information.
This summarised view enables prioritising threats that require exploring additional information such as the tweets' links.

Twitter was chosen for two main reasons.
First, Twitter is well-recognised as a relevant source of short notices (almost in real-time) about web activity and occurring events~\cite{howpeopleusetwitter}.
Previous research shows this is also true for cybersecurity~\cite{sabottke2015vulnerability,Campiolo13,mcneil2013pace}.
In fact, the most important cybersecurity news feeds are present in Twitter (\eg{}, NVD, ExploitDB, CVE, Security Focus, Nessus), making it a hub for all these sources.
Second, the limited size of a tweet makes it simple to process through general-purpose machine learning approaches, which enable low error levels across multiple domains of application.
Furthermore, although short, tweets provide enough elements to categorise their content, as well as links for more detailed material.

Most previous work to gather cybersecurity OSINT information focuses on the filtering and classification process~\cite{sabottke2015vulnerability,mittal2016cybertwitter,ritter2015weakly,mathews2012collaborative,le2017sonar}.
Beyond that, few works extract information from unstructured text (including tweets)~\cite{liao2016acing,zhu2016featuresmith,trabelsi2015mining,dionisio2019cyberthreat}.
However, to the best of our knowledge, no previous work addresses the timely summarisation of a cyberthreat Twitter stream, thereby providing an end-to-end approach for monitoring the current threat landscape.

The standard technique for aggregating related data is clustering~\cite{aggarwal2015data}.
In a streaming context, a stream clustering algorithm becomes necessary~\cite{silva2013data}. 
Existing algorithms (\eg{},~\cite{zhang1996birch,guha2000clustering,zhou2008tracking}) have two shortcomings for our context: they require \emph{a priori} definition of the target number of clusters and they discard outliers.
However, when processing a cybersecurity news feed, the number of active threats under discussion is unknown in advance and outliers cannot be discarded as they are likely to represent new threats.

\toolname{} adapts a well-known stream clustering algorithm to overcome these limitations, by detecting whether tweets refer \emph{new threats} or \emph{updates} to previously known ones, thus becoming appropriate for maintaining a continuous up-to-date summary of current threats observed.
Finally, to close the pipeline from Twitter to the SOC tools, \toolname{} produces IoCs from the clustered OSINT, making it integrable with various SIEMs (\eg{}, IBM QRadar~\cite{qradar}) and threat intelligence/sharing tools (\eg{}, MISP~\cite{misp}).

A quantitative evaluation considering all tweets from 80 accounts over more than 8 months (over 195.000 tweets), shows that \toolname{} finds the majority of security-related tweets concerning an example IT infrastructure (true positive rate above 90\%), incorrectly selects a small number of tweets as relevant (false positive rate under 10\%), and summarises the results to very few IoCs per day.
When compared to a naive text-filtering approach (as employed by most threat intelligence systems used in practice), it decreases the number of tweets presented by approximately 80\%, with the number of summarised IoCs being only 21\% of the tweets classified as relevant.
This volume of data can either be inspected manually or processed by a SIEM as OSINT-generated events.
Further, a qualitative analysis of the largest 65 clusters generated by \toolname{} revealed two paramount findings.
Firstly, 43\% of the IoCs describe high-impact security alerts (CVSS $\ge 7.0$), and for half of these, the tweet publication preceded the vulnerability publication on the National Vulnerability Database (NVD) by eight days (on average).
Secondly, 70\% of the analysed clusters provided serviceable intelligence, including exploits whose vulnerabilities were not matched to NVD entries.
In summary, our contributions are:

\begin{enumerate}
\item An end-to-end streaming threat monitor architecture for collecting, classifying, and clustering tweets related to a specified infrastructure (Section~\ref{sec::methodology});
\item A novel application strategy and adaptation of well-known clustering techniques to the context of cybersecurity threat awareness (Section~\ref{sec::methodology_clustering});
\item A detailed system evaluation using three real-world datasets and a qualitative analysis of the security alerts generated thereof (Section~\ref{sec::results});
\item Methods for generating MISP-compatible IoCs from tweets that enable the integration of \toolname{} into SOC operation (Sections~\ref{sec::ioc_generation} and \ref{sec::pragmatics}).
\end{enumerate}

\section{Related Work}
\label{sec::related-work}

In the following, we briefly review the previous work related to \toolname{}: processing tweets for cybersecurity, threat intelligence tools, and stream clustering algorithms.

\paragraph{Twitter for cybersecurity.}
Several works aim to find cybersecurity OSINT about a given IT infrastructure.
These rely on a keyword set to govern the selection of tweets, thereby picking only the potentially relevant content.
Mittal \etal{}~\cite{mittal2016cybertwitter} use a knowledge base created from security concepts to evaluate if a tweet is relevant for cybersecurity.
Similarly, Le Sceller \etal{}~\cite{le2017sonar} designed a framework that collects tweets on a keyword basis and is capable of extending the keyword set automatically.
Ritter \etal{}~\cite{ritter2015weakly} search Twitter for occurrences of three specific \emph{topics}: DoS attacks, data breaches, and account hijacking.
Trabelsi \etal{}~\cite{trabelsi2015mining} cluster tweets by subject. Threats not referred by NVD are considered novel and handled like zero-day vulnerabilities.
Dionísio \etal{}~\cite{dionisio2019cyberthreat} used deep learning techniques to detect and extract security-related information from tweets.
Sabottke \etal{}~\cite{sabottke2015vulnerability} show that information about exploits are published on Twitter two days before they are included in NVD (on average).
None of these works provide an end-to-end solution for online threat monitoring, mainly because they focus on detection, overlooking summarisation and SOC integration.

\begin{figure*}[ht]
    \centering
    \includegraphics[width=\textwidth]{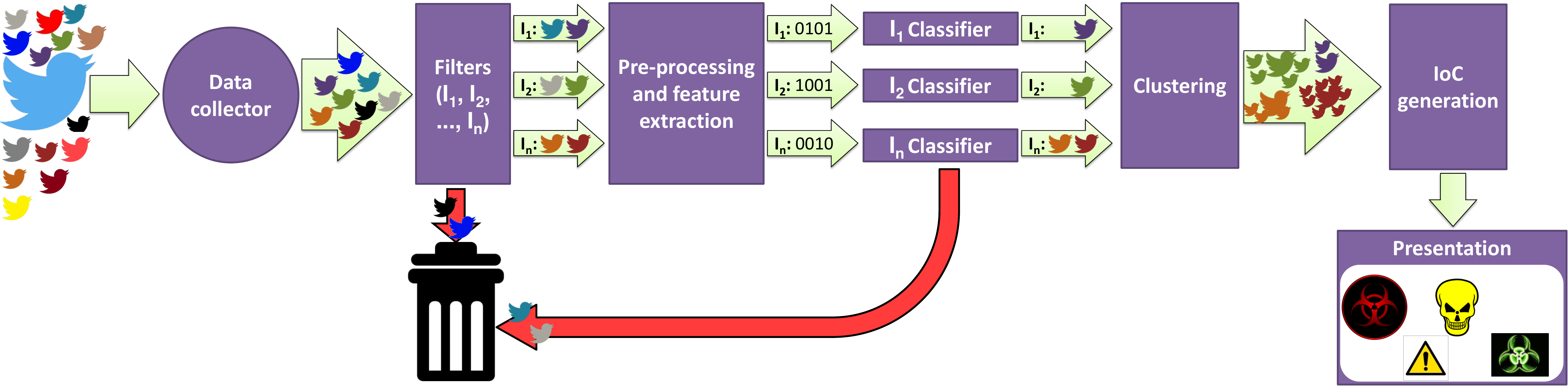}
    \caption{\toolname{}'s architecture. Collected tweets pass through the various stages and those classified as relevant are aggregated, transformed in IoCs, and delivered to SOC analysts.}
    
    \label{fig::architecture}
\end{figure*}

\paragraph{Threat Intelligence Tools.}
Research-oriented work focus on gathering OSINT and transforming it into machine-readable IoCs for feeding Intrusion Detection Systems (IDS), anti-viruses, or other tools.
Mathews \etal{}~\cite{mathews2012collaborative} employ traditional (\eg, logs) and non-traditional (\eg, forums, blog posts) data sources to create an ontology that infers the legitimacy of traffic flows, feeding an IDS with the results.
Liao \etal{}~\cite{liao2016acing} developed a framework for extracting IoCs from technical literature, enabling high recall of the methodology.
In a different work, Zhu \etal{}~\cite{zhu2016featuresmith} present a system that processes the scientific literature studying Android malware and extracts features describing the attacks to create a malware detector.
The objective of these works is to extract machine-readable information from OSINT, which is different from our goal.

Besides the research-oriented efforts to include OSINT in protection systems, off-the-shelf tools are able to collect and deliver OSINT-based threat intelligence.
SpiderFoot~\cite{spiderfoot} is an OSINT automation tool that uses multiple sources (\eg, Bitcoin addresses, Twitter) for three main purposes: target reconnaissance, assess an organisation's exposure on the Internet, and OSINT collection for security purposes.
IntelMQ~\cite{intelmq} is an open-source system for collecting and processing security-related OSINT feeds designed for organising data coming from various sources.
It employs an ontology for data harmonisation and converts all events into a uniform \texttt{json} format.
MISP~\cite{misp} is a threat intelligence platform designed for sharing and correlating IoCs.
It receives many types of threat inputs and exports its data into other MISP instances or threat intelligence tools.
Generally speaking, these tools do not employ any advanced processing capability for filtering and matching threats, resorting only to keyword-based string comparisons.

\paragraph{Stream clustering.}
With the few exceptions discussed bellow, most stream clustering algorithms require the target number of clusters ($k$) to be defined as a parameter and discard elements that do not fit the clusters (outliers)~\cite{silva2013data}.
Feng \etal{}~\cite{feng2015streamcube} cluster only the tweets' hashtags, using text similarity to adapt the number of clusters to the collected data.
However, this algorithm would potentially miss important information in the security field, as the clustering would not consider the full tweet text, only hashtags.
Saki \etal{}~\cite{saki2016online} use a density-based clustering approach, therefore avoiding the definition of $k$.
However, their technique discards outliers, which could lead to missing important emerging threats.
Shou \etal{}~\cite{shou2013sumblr} algorithm allows the value of $k$ to vary up to an upper limit, but its outlier detection mechanism discards topics that do not gain traction, ignoring possibly important threats that remain unknown for long periods of time. 

\section{\toolname{} Pipeline}
\label{sec::methodology}

Figure~\ref{fig::architecture} presents \toolname{}'s architecture and data processing stages---tweet gathering, filtering, feature extraction, classification, clustering, and IoC generation---described next.

\paragraph{Data Collection.}
The data collector module requires a set of accounts, from which it will collect every posted tweet using Twitter's stream API---an approach already found in the literature~\cite{sapienza2018discover}.
These accounts can be from security analysts and organisations, vendors, hackers, researchers, among others.
They are chosen considering the likelihood of users tweeting about the security of elements belonging to the monitored IT infrastructure.
Since usually security analysts already follow OSINT sources and Twitter accounts, it is just a matter of providing these sources to \toolname{}.

Simply collecting tweets by keywords is a method likely to retrieve large amounts of irrelevant information.
For instance, tweets with the word ``windows'' include all Windows-related topics (the OS) and all tweets referring glass windows.
By collecting tweets only from selected security-related accounts, a more substantial fraction of tweets is related to cybersecurity.

\paragraph{Filtering.}
Despite the account-based collection approach, most likely the collected data will include tweets unrelated to the infrastructure under the analyst's care.
These have to be dropped by a filter.
The filtering approach assumes that a tweet referring a threat to a particular IT infrastructure asset has to mention that asset.
Therefore, a second input is required: a set of keywords describing the assets of the monitored IT infrastructure.
Only tweets that include at least one of the keywords will pass the filter.
Keywords further restrict the scope of the security events, hence decreasing the number of irrelevant tweets beyond the filter.

To maximise the effectiveness of \toolname{}, the keywords defining the monitored assets must be as complete and specific as possible.
For example, if the analyst is in charge of securing a Linux cluster running virtual machines to serve a web service with a database, the keyword set could be \texttt{\{linux, ssh, virtualbox, vbox, mysql, apache, php\}}.

\paragraph{Pre-processing and Feature Extraction}
Pre-processing normalises the tweet representation.
First, all characters are converted to lower case, and stopwords and hyperlinks are removed---the latter are shortened URLs that provide little information.
Numbers, dots, and hyphens are replaced by their textual representation (\eg, ``2'' to ``two''), as these are relevant to distinguish software versions (\eg, Mozilla Firefox 4.5.1-2).
Finally, all non \texttt{[a-z]} characters are removed.
For instance, after pre-processing, the tweet ``\emph{\#Oracle \#Linux 6 / 7 : Unbreakable Enterprise kernel (ELSA-2016-3573) https://t.co/vLTel8NodG}'' becomes ``\emph{oracle linux six seven unbreakable enterprise kernel elsa hyphen two thousand and sixteen hyphen three thousand five hundred and seventy three}''.
The original tweets are stored for presentation.

The tweets must be converted to a numerical format to become suitable for supervised learning classification techniques.
This work uses the well-known Term Frequency - Inverse Document Frequency (TF-IDF) method~\cite{leskovec2014mining}.
TF-IDF computes weights to words (features) based on their occurrence frequency in each document and on the group of documents considered.
The weight of a word increases with its frequency of occurrence in a single document but is scaled down by the frequency of occurrence in all documents.
By mapping each consecutive word token to a corresponding vector position, tweets are converted to a constant size, zero-padded, TF-IDF numeric vector.
Finally, to limit the size of the vector we employ the hashing trick technique~\cite{weinberger2009feature}.

\paragraph{Classification.}
For the classification of tweets according to their security relevance, two classifiers have been explored: Support Vector Machines (SVM)~\cite{cortes1995support} and Multi-Layer Perceptron (MLP) Neural Networks (NN)~\cite{rosenblatt1958perceptron,rumelhart1985learning}.
The SVM is a broadly-used classifier achieving good results across a multitude of application domains.
We consider the SVM implementation available in the Apache Spark' Machine Learning library (MLlib)~\cite{spark}, which employs a linear kernel, thereby assuming the input vectors are linearly separable.

Since MLlib does not provide a non-linear SVM kernel, MLlib's MLP NN implementation was considered to account for the assumption that input vectors may not be linearly separable.
The MLP is a well-established and frequently used NN architecture that has a long track record of good and consistent results over a vast number of classification tasks.

\paragraph{Clustering.}
\toolname{} uses clustering to aggregate similar tweets in the news feed stream.
The Clustream algorithm~\cite{aggarwal2003framework} was chosen as the basis for this pipeline stage as its structure and characteristics were closest to our requirements.
However, it required adaptation to \toolname{}'s context to achieve threat aggregation as described in the next section.

\begin{figure}[t]
\scriptsize
\centering
    \includegraphics[width=1.0\columnwidth]{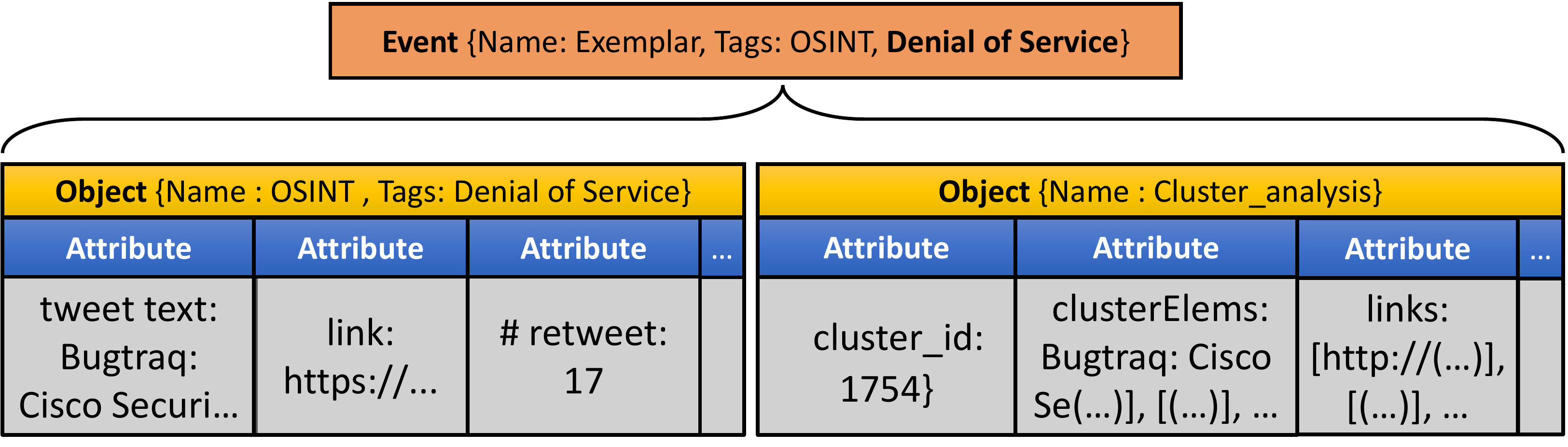}
    \frame{\includegraphics[width=1.0\columnwidth]{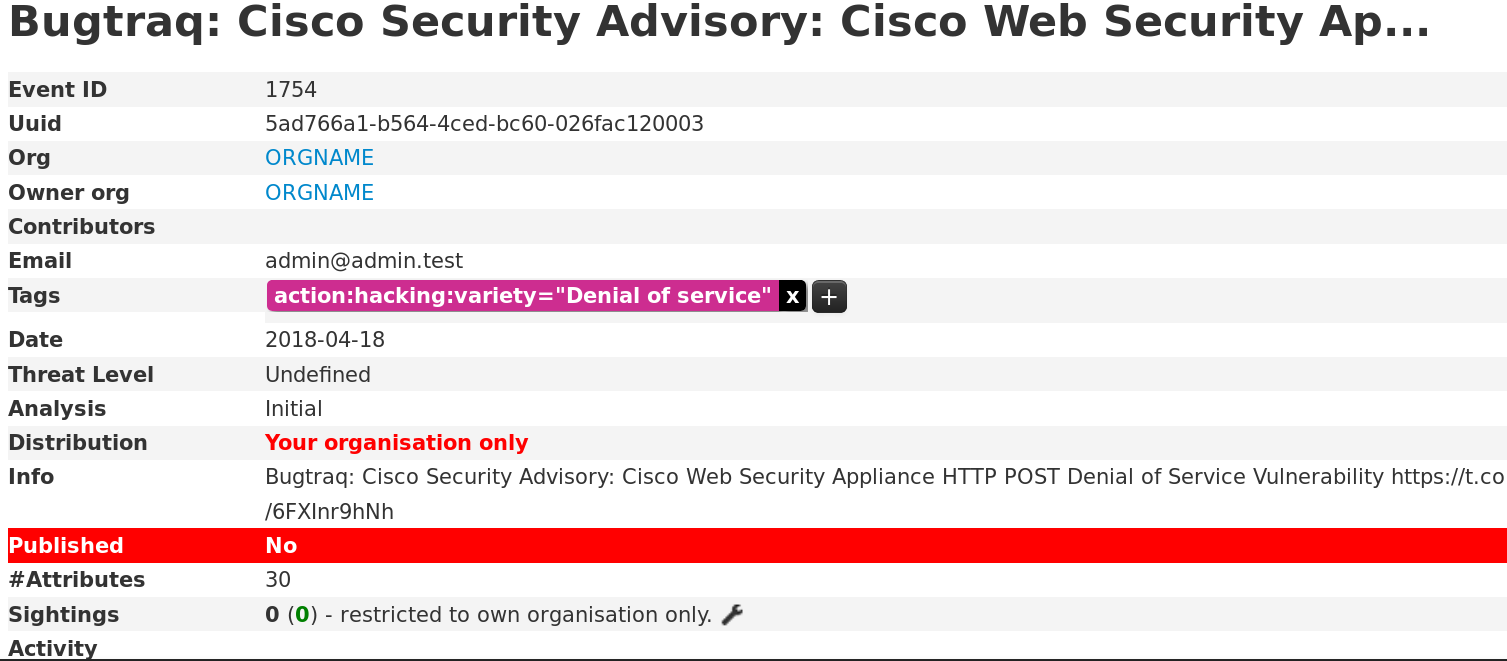}}
    \caption{Representation of a cluster into the MISP taxonomy~\cite{misp-format} and an OSINT-generated event in MISP.}
    \label{fig::misp}
\end{figure}

\paragraph{MISP compatible IoC Generation.}
\label{sec::ioc_generation}
After the clustering phase, the clusters of tweets are transformed into the IoC format to allow their inclusion in SIEMs or threat intelligence platforms.
There are several standards for sharing IoCs, such as STIX~\cite{stix} or MISP~\cite{misp-format}.
The format must be extensible and adaptable as tweets are unstructured and contain unpredictable content.
For these reasons, the MISP format has been selected to generate IoCs.
Moreover, it can be easily converted into other standard formats like STIX.

We use a combination of MISP items to generate the IoC.
One MISP \textit{Event} is composed of two \textit{Objects} containing security indicators called \textit{Attributes}: one describing the content of the exemplar tweet (the cluster centroid); the other representing the cluster of tweets.
Events are classified using tags, added according to a set of threat categories related to existing taxonomies: ENISA and VERIS for cyberthreats~\cite{misp-taxonomies}.
The OSINT tag is added to emphasise the automatic creation based on tweets.
The classification is achieved by using regular expressions to match taxonomy elements in the exemplar's message, generating one tag for each match.

Figure~\ref{fig::misp} depicts the taxonomy employed to represent IoCs in MISP (top of the figure).
The exemplar tweet is the core of the IoC, while its cluster is an extra element to increase informativeness.
The bottom of the figure shows a MISP Event generated from a cluster and its exemplar (the example cluster shown in Table~\ref{table::cluster_example}).
The OSINT object contains extracted information from the exemplar such as the tweet's message, any links therein, and the \textit{Cluster\_Analysis} object contains the remainder cluster data.
A simple classification was applied: the OSINT tag marks the event as created from tweets, and the ``Denial of Service'' tag (from VERIS) classifies the threat.

\begin{table}[t]
    \centering
    \caption{An example of a cluster and its \textit{exemplar} (in Bold).}
    \scriptsize
    \begin{tabular}{|p{0.95\columnwidth}|}
    \hline
\textbf{Bugtraq: Cisco Security Advisory: Cisco Web Security Appliance HTTP POST Denial of Service Vulnerability https://t.co/6FXInr9hNh} \\\hline
Bugtraq: Cisco Security Advisory: Cisco Web Security Appliance HTTP POST Denial of Service Vulnerability https://t.co/6FXInr9hNh \\\hline
Bugtraq: Cisco Security Advisory: Cisco Web Security Appliance HTTP Length Denial of Service Vulnerability https://t.co/TgU0T9vlZt \#bugtraq \\\hline
Bugtraq: Cisco Security Advisory: Cisco Web Security Appliance HTTP POST Denial of Service Vulnerability https://t.co/feZlTxQKVC \#bugtraq \\\hline
\#cybersecurity Bugtraq: Cisco Security Advisory: Cisco Web Security Appliance HTTP POST Denial of Service https://t.co/XUUctUnQ8F \#infosec \\\hline
\#vulnerability \#security : Bugtraq: Cisco Security Advisory: Cisco Web Security Appliance HTTP POST Denial of Serv https://t.co/9bW0ls00kx \\\hline
\#internet \#security: Cisco Web Security Appliance HTTP POST Denial of Service Vulnerability https://t.co/cXQUTWUBbD \\\hline
    \end{tabular}
    \label{table::cluster_example}
\end{table}

\section{Tweet stream clustering}
\label{sec::methodology_clustering}

Since Twitter users can tweet or retweet about the same subject, \toolname{} is expected to collect many similar tweets.
Thus, to cover information about the IT infrastructure, the analyst would have to manually inspect a large amount of redundant data for each threat.

To alleviate this burden, clustering is used to group similar tweets classified as relevant for the protection of the IT infrastructure.
Ideally, the information collected about a specific threat gets aggregated in one cluster, from which a single representative tweet---the \emph{exemplar}---is presented to the analyst.
By clustering the stream of relevant tweets, distinct active threats are summarised in a set of clusters and updated as more tweets are collected.
It is through this mechanism that \toolname{} can create an active threat monitor outlining the current threat landscape, \ie{}, the current threats that potentially require more immediate attention from SOC analysts. 

\subsection{Data stream aggregation challenges}

Clustering is commonly applied in batch, as an exploratory data technique where a static data set is clustered into $k$ groups~\cite{aggarwal2015data}.
The number of clusters, $k$, is either defined \emph{a priori} or estimated to satisfy performance metrics~\cite{aggarwal2015data}.
In a dynamic setting such as \toolname{}'s streaming context, defining $k$ beforehand is not possible, as the number of threats being discussed at a given time is unknown.
If at any moment \toolname{} was processing $t$ threats and clustering was set to find $k\neq t$ clusters, the result would contain clusters including unrelated threats, various clusters related to the same threat, or both cases.
\emph{Therefore, \toolname{} requires a clustering algorithm able to adapt $k$ over time.}

Furthermore, an essential feature of most stream clustering algorithms is the ability to detect and remove outliers that may disrupt the quality of the clustering.
In the security context, performing outlier removal could prevent the discovery of emerging threats.
Moreover, all tweets reaching \toolname{}'s clustering stage were classified as relevant, and should not be discarded.
\emph{Therefore, \toolname{} requires a clustering algorithm capable of maintaining performance indicators (\eg{}, intra and inter-cluster cohesion) without removing outliers}.

\subsection{DynamicClustream}

The lack of solutions that fit the requirements of threat intelligence tools (see Section~\ref{sec::related-work}), motivated us to adapt the Clustream~\cite{aggarwal2003framework} algorithm for \toolname{}, thus creating the DynamicClustream.
The Clustream algorithm clusters a data stream in two phases.
The online phase performs a simple and efficient clustering of the inbound stream by keeping only a summary of the data collected, thus abiding to the speed requirements of a data stream~\cite{aggarwal2015data}.
The offline phase is performed in background to provide a more complete analysis of the collected data through a more effective and computationally demanding clustering algorithm.
Clustream includes an outlier detection mechanism that excludes data points unfit for any of the existing clusters by analysing the distance from that point to all clusters.
A decision is only taken once it becomes clear if a data point is an element of a new trend or an isolated occurrence.
The components that distinguish DynamicClustream from Clustream are detailed in the following.

\paragraph{High-level Overview.}
Assume there is always a global cluster state $S$, defined as a set of sets, describing the clusters formed from a previously processed time-window of tweets.
When a new tweet $t$ is received, the online clustering component attempts to place $t$ in one of $S$'s clusters.
If a direct placement is not possible, the offline clustering component is triggered to compute a new clean cluster state considering the tweets in the clusters of $S$ plus $t$.

Once a new cluster state $S$ is in place, a final step is taken to obtain each cluster's \emph{exemplar} tweet, \ie{}, the tweet representing the cluster, that will be shown to the analyst.
The exemplar tweet is selected by choosing the tweet with the smallest Euclidean distance to the centroid of the cluster.
An example of a generated cluster (and its exemplar) appears in Table~\ref{table::cluster_example}.
The online and offline components of DynamicClustream are presented in Algorithm~\ref{alg::clustering}, with locking details for ensuring atomic updates on $S$ omitted for better readability.

\paragraph{Online clustering component.}
The online clustering component uses a lightweight approach to assign a new tweet $t$ to the current clustering state $S$.
To do so, the membership of $t$ is tested in all clusters (line 3) by employing the WTS cohesion measure (introduced below).
This is done by adding $t$ to each cluster $C_i \in S$ and calculating the corresponding WTS value.
$t$ belongs to $C_i$ when WTS is above a certain threshold $\tau$.
If $t$ does not fit in one of the existing clusters, a new cluster solely containing $t$  is created (lines 4--5).
If $t$ belongs to a single cluster, it is added to that cluster (lines 6--7).
When $t$ fits more than one cluster, it is added (temporarily) to the cluster with the highest membership rate, and the offline clustering is scheduled (lines 9--10).

In \toolname{}'s application scenario it makes no sense to remove outliers.
Instead, when new tweets do not belong to $S$, we treat them as the onset of a threat by adding new clusters with a single element which in time may receive additional tweets.
This outlier processing mechanism allows adapting the number of clusters, $k$, to the novelty in the dataflow.
Furthermore, it is through the online component of DynamicClustream that the active threat monitor is implemented: the system categorises new tweets as \emph{new threats} or as \emph{updates} to known ones, thus maintaining an updated threat summary about an IT infrastructure.

\begin{algorithm}[t]
\caption[caption]{DynamicClustream online and offline clustering.}
\footnotesize
\label{alg::clustering}
\DontPrintSemicolon
$S \gets \emptyset$ \tcp*{global cluster state}
\vspace{1mm}
\SetKwFunction{OnClu}{OnlineClustering}
\SetKwProg{Fn}{Function}{:}{}
\Fn{\OnClu{$t$}}{
        $i \gets \mathsf{GetNumHits}(S, t)$\;
        \If{$i = 0$}{
            $\mathsf{AddNewCluster}(S, t)$\;
        }\ElseIf{$i = 1$}{
            $\mathsf{UpdateCluster}(S, t)$\;
        }\Else(\tcp*[f]{needs offline clustering}){
            $\mathsf{PlaceInClosestCluster}(S, t)$\;
            \textbf{schedule} $\mathsf{OfflineClustering}(S)$ 
        }
}

\SetKwFunction{rclu}{OfflineClustering}
\SetKwProg{Fn}{Function}{:}{}
\Fn{\rclu{SavedState}}{
        $\mathit{T} \gets \mathsf{Flatten}(SavedState)$\;
        $\varepsilon^* \gets +\infty$; $k \gets 2$; $Clusters \gets \emptyset$\;
        $S^* \gets \emptyset$\;
        \While{$\mathit{T} \neq \emptyset$}{
            \Do{$\varepsilon = \varepsilon^*$ \textbf{and} $k < |\mathit{T}|$}{
                 $Clusters, \varepsilon \gets \mathsf{KMeansClustering}(\mathit{T}, k)$\;
                 \If{$\varepsilon < \varepsilon^*$}{
                     $\varepsilon^* \gets \varepsilon$\;
                    $k \gets k+1$\;
                 }
            }
            \ForAll {$C \in Clusters $}{
                 \If{$\mathsf{WTS}(\mathit{C}) \geq \tau$}{
                     $S^* \gets S^* \cup \{C\}$\;
                     $\mathit{T} \gets \mathit{T} \setminus C$\;
                 }
            }
        }
        $S \gets MergeClusterState(S^*,\mathsf{Flatten}(S) \setminus \mathsf{Flatten}(S^*))$\;
}
\end{algorithm}

\paragraph{Cohesion Measure.}
Cluster cohesion and cluster separation are concepts used to assess the validity of a partition generated by a clustering algorithm~\cite{Arbelaitz13cvi}, which in most cases have a purely geometric interpretation.
In \toolname{}, cohesion is based on the similarity of tweets within a cluster and not on a geometric measure such as the distance to the cluster centroid, thus defining a context-based cluster validation approach, argued to be more effective~\cite{guyon2009clustering}.

To reinforce the one-to-one relation between clusters and threats, the cohesion measure must detect clusters whose tweets refer to the same threat.
Assuming that a threat is expressed by a minimum number of words appearing in all tweets, the proposed cohesion measure---named \emph{Within-cluster Threat Similarity} (WTS)---is defined as $\frac{\omega}{w_m}$, where $\omega$ is the number of words shared by all the cluster's tweets and $w_m$ is the number of words of the smallest tweet in the cluster.
WTS is 0 if no words are shared by the tweets of a cluster, and 1 when all tweets share the words of the smallest tweet in the cluster.
It assumes that if all cluster tweets share a sufficiently large number of words, then they mention the same threat.

The degree of separation of two clusters $C_i$ and $C_j$ is measured by the Jaccard index~\cite{zaki2014data}.
It is determined as $J=\frac{|C_i \cap C_j|}{|C_i \cup C_j|}$, corresponding to the ratio between the number of common words to $C_i$ and $C_j$ and the number of unique words of $C_i$ and $C_j$.
The lower its value, the more separated the clusters are.

\paragraph{Offline clustering component.}
The offline component applies the \kmeans{} clustering algorithm~\cite{mcqueen67kmeans} repeatedly to provide more robust clusters.
\kmeans{} is a widely used algorithm that has provided good efficiency and empirical success over the last 50 years~\cite{jain10clustering}.
However, it is commonly employed for exploratory data analysis, not for automatic text summarisation.

The \kmeans{} algorithm requires the specification of the number of clusters, $k$, which is unknown in this case.
At a given time we do not know how many potential threats to our infrastructure are being discussed.
Therefore, we defined a novel strategy to find the so-called elbow point~\cite{tibshirani2001estimating}, \ie, the point beyond which by increasing $k$ there is no significant improvement in the clusters' Sum of Squared Errors (SSE).
This procedure automatically determines $k$, thus avoiding the specification of a threshold to find the elbow point or the visual inspection of the within-class-variance versus $k$ graph.

\textit{\kmeans{} application strategy:}
Starting at $k_1=2$, a \kmeans{} model is trained for each successive $k_i=i+1$ number of clusters, which produces a corresponding SSE, denoted by $\varepsilon_i$.
As the initial cluster centres are randomly chosen, there is a given variance $\sigma_i$ associated to $\varepsilon_i$.
As we keep increasing $k_i$, we expect $\varepsilon_i$ to decrease up to the point where the magnitudes of $\varepsilon_i$ and $\sigma_i$ become of the same order.
At this point $\varepsilon_{i+1}-\varepsilon_i$ might become zero or even negative, indicating that there is no significant SSE improvement in increasing $k_i$.
Therefore, the iteration is stopped when the error ($\varepsilon$) stops decreasing or (the limit case where) the number of clusters corresponds to the number of tweets to be clustered, and $k_i$ is selected as the number of clusters (lines 16--21).

By testing this approach, we found that small clusters had only very similar tweets, but other large clusters contained unrelated tweets.
The cause might be two-fold: (1) \kmeans{} assumes spherical clusters that it tends to produce equally sized, which might not be adequate; and (2) the strategy to find $k$ is not guaranteed to find the \emph{best} $k$.
To overcome this limitation, we use the WTS cohesion measure to quantify how closely related the tweets in a cluster are, and implement a \emph{re-clustering method} that splits these clusters into smaller ones with related tweets.
If WTS $\geq \tau$ (a specified threshold), indicating high cohesion, it enables the validation of clusters as \emph{final}.

\textit{Re-clustering method:}
All tweets of non-final clusters are gathered (line 22--25) and re-clustered (lines 16--21) using \kmeans{} to allow similar tweets to be grouped.
Then, the new clusters generated are again tested using their WTS, and the process is repeated for the non-final clusters.
Eventually, all clusters are considered final, ideally each related to a single threat, and $S^*$ is merged with $S$ (line 26), i.e.,  $S^*$ is updated with the tweets received since the algorithm started by executing a procedure similar to lines 3--10.

\textit{Offline clustering scheduling:}
At any time, there may be only one instance of the offline component in execution. 
Since multiple tweets received in a short time interval may trigger offline clustering, we employ the schedule keyword (line 10) to avoid overlapping executions.
The idea is that each call to \textbf{schedule} $\mathsf{OfflineClustering}()$ notifies the system that offline clustering is required after this point, and saves the current cluster state for its next execution.
Once the algorithm is started again (using the latest saved state), it process all tweets pending in $S$ (line 12).

\paragraph{Time-window model.}
To fully adapt Clustream to our context we also changed the clustering ageing model used to remove clusters.
This model is necessary to complete the adaptation of the cluster state to the data stream flow.

Clustream's window model is global in the sense that all data points are aged and removed using the same rule.
However, this methodology does not fit \toolname{} application domain, as different cybersecurity topics have different lifetimes.
For example, news about an update are expected to last a few days, while advances about an active threat may continue for a month or more.
Thus, in the cybersecurity field it makes more sense to adopt a local window model, monitoring ageing \emph{by cluster} (by threat).
As a consequence, whole clusters rather than single points should be removed in forthcoming clustering states.

In DynamicClustream a cluster $C_i$ is removed from the cluster state $S$ if it has been stale for a period of time longer than $\theta$, \ie{}, if $\theta$ time passes without $C_i$ receiving a new data point.
In this way, topics that no longer receive traction are stowed away, \emph{while active topics retain all their elements, regardless of the time passed, which may be crucial for understanding the evolution of a threat}.

\section{SOC integration}
\label{sec::pragmatics}

An essential aspect of threat intelligence tools such as \toolname{} is the integration in a SOC.
In the following, we describe practical issues related to this integration.

\paragraph{Twitter as OSINT.}
When using Twitter as a cybersecurity information source, it is important to consider what would happen if some of the monitored accounts fell under the control of the adversary.
In a nutshell, two things can happen~\cite{sabottke2015vulnerability}: (1) the adversary may not tweet about the threats he is interested in exploiting using the accounts he controls; or (2) the adversary may create tweets with false threats, to make SOC analysts waste their time in solving potential non-existent problems.
Both attacks should not be a significant problem as long as the amount of accounts controlled by the adversary is relatively small, and the analysts take into account the reputation of the accounts monitored by the system.

\paragraph{Training the system.}
Our approach requires the creation of labelled datasets for training the classifiers.
To do that, the SOC analysts need first to configure the keywords defining the infrastructure.
A second configuration step is to define the Twitter accounts that will be monitored.

After those two steps, the system should present all filtered tweets as if they are important, and a button for the analyst to mark a tweet as ``irrelevant''.\footnote{The ``irrelevant'' button must always be available, even when the system is not being trained, in order to collect wrongly classified tweets for future retraining.}
Notice that, to avoid bias, it is relevant to inform the analysts that the system is under training.
When enough positively-labelled tweets are collected, the classifiers can be trained in background and then placed in operation.

It is expected that the classifier's performance decreases with time, as the operational data gets progressively different from the training data.
To maintain the utility of the classifiers in use, it is essential to minimise this effect.
Incremental learning is a technique that can be used for this purpose, where the classifier's model is continuously trained with new labelled examples~\cite{geng2015incremental}.
By training the model with the latest events, it is continuously adapted to changes in input format (in this case, changes in tweet format or language).

Another possibility is to replace the model with a new model trained with only the latest data, \eg, the last three months of tweets.
This way the model is periodically adapted to the current threat landscape, so that old data will not impact the classifier's quality.

\paragraph{Changing keywords and monitored accounts.}
Adding or removing keywords from the datasets require retraining the classifier.
Removing a keyword requires removing the tweets that were filtered by this keyword and retrain the model without them.
To add a keyword, one needs first to complement the existing labelled dataset (in the same way as described before) with tweets related to the new keyword, and then retrain the model with the reformulated data set.
Changing the set of monitored Twitter accounts is not a burden for the system since the structure of threat descriptions is expected to be similar across all security accounts.
The datasets employed in our experimental evaluation consider this possibility.

\section{Experimental Setup}
\label{sec::setup}

This section describes the experimental work carried out to validate \toolname{}.
All code is written in Scala and deployed on the Apache Spark Framework~\cite{spark}.
We chose Spark as its data-structures are scalable and designed for large datasets.
Also, Spark includes a scalable machine learning library called MLlib, used to implement all ML algorithms employed in this paper.

\paragraph{Infrastructure Definition.}
We used a hypothetical IT infrastructure to set \toolname{}'s filter during its experimental evaluation.
This infrastructure (presented in Table~\ref{table::infrastructures}) is composed of software elements typically found in the IT world, such as the most common browsers and operating systems.

\begin{table}[t]
    \centering
    \scriptsize
    \caption{The hypothetical infrastructure designed for tweet collection and filtering.}
    \begin{tabularx}{.95\columnwidth}{|Y|}
        \hline
        oracle, cisco, internet explorer, google chrome, chrome, firefox, microsoft edge, edge, wordpress, joomla, wp, microsoft windows, ms, linux, operating system, operating systems\\
        \hline
    \end{tabularx}
    \label{table::infrastructures}
\end{table}
\begin{table}[t]
    \centering
    \caption{Datasets collection and labeling details.}
    \scriptsize
    \begin{tabular}{|c|c|c|c|c|c|c|}
        \hline
        \textbf{Dataset:} & \multicolumn{2}{c|}{\texttt{D1}} & \multicolumn{2}{c|}{\texttt{D2}} & \multicolumn{2}{c|}{\texttt{D3}} \\
        \hline \hline
        Time period & \multicolumn{2}{c|}{01/11/2015} & \multicolumn{2}{c|}{01/04/2016} & \multicolumn{2}{c|}{15/05/2016}\\
         (from/to)  & \multicolumn{2}{c|}{01/04/2016} & \multicolumn{2}{c|}{15/05/2016} & \multicolumn{2}{c|}{10/07/2016}\\
        \hline
        Account sets & \multicolumn{2}{c|}{\texttt{S1}} & \multicolumn{4}{c|}{\texttt{S1}, \texttt{S2}}\\
        \hline
        Total tweets collected & \multicolumn{2}{c|}{71024} & \multicolumn{2}{c|}{57579} & \multicolumn{2}{c|}{66608}\\
        \hline
        \multirow{2}{*}{Class distribution} & Pos. & Neg. & Pos. & Neg. & Pos. & Neg. \\
        \cline{2-7}
       ~ & 1697 & 2008 & 536 & 4292 & 1680 & 2153 \\
        \hline
    \end{tabular}
    \label{table::datasets}
\end{table}

\paragraph{Tweet Collection and Labelling.}
We collected three datasets during three periods of time.
Table~\ref{table::datasets} presents their collection periods, the sets of accounts used, and the number of tweets.
After being collected and filtered using the keywords in Table~\ref{table::infrastructures}, each tweet was manually labelled as positive (the tweet mentions a threat to a given part of the IT infrastructure) or negative, thus creating labelled datasets suitable for supervised learning.

Two sets of accounts, \texttt{S1} and \texttt{S2}, were used for tweet collection, as shown in the third row of Table~\ref{table::datasets}.
The accounts are listed in Table~\ref{table::all_accounts}.

\begin{table}[t]
    \centering
    \caption{Sets of accounts used to create the datasets.}
    \scriptsize
    \begin{tabular}{|p{0.95\linewidth}|}
    \hline
    \textbf{\texttt{S1} Accounts:} 
    inj3ct0r, TrustedSec, Anomali, briankrebs, Secunia, exploitdb, alienvault, slashdot, dstrom, Info\_Sec\_Buzz, vuln\_lab, threatintel, dangoodin001, ivspiridonov, ThreatFeed, pikisec, SANSInstitute, johullrich, drericcole, F1r3h4nd, MaldicoreAlerts, USCERT\_gov, gcluley, hal\_pomeran, SecurityWeek, SecurityNewsbot, sans\_isc, e\_kaspersky \\
    \hline
    \textbf{\texttt{S2} Accounts:} 
    TenableSecurity, securitywatch, securityaffairs, zer0element, notsosecure, CyberExaminer, SCMagazine, DMBisson, lennyzeltser, IT\_securitynews, teamcymru, WordPress, MicrosoftEdge, JoomlaTips, sjzaib, SecurityMagnate, Cisco, Dell, linuxtoday, securityninja, cyberopsy, OWASP\_Java, \_WPScan\_, d\_plusk, threatpost, Rootsector, Microsoft, linuxfoundation, ChidoDike, Sec\_Cyber, ptracesecurity, msftsecurity, LinuxSec, hack3rsca, CiscoSecurity, NytroRST, joomla, Windows, crackerhacker00, fstenv, HPE\_Security, googlechrome, wordpressdotcom, packet\_storm, RokaSecurity, Oracle, firefox, wpbeginner, YoKoAcc, SecurityCrap, jasonlam\_sec, threatmeter \\
    \hline
    \end{tabular}
    \label{table::all_accounts}
\end{table}

\begin{table}[t]
    \centering
    \caption{The words used in the Logstash filter.}
    \scriptsize
    \begin{tabular}{|p{0.95\linewidth}|}
    \hline
access, acl, admin, advisory, allow, arbitrary, aslr, assurance, attack, auth, buffer, bug, bypass, certificate, code, command, corruption, csrf, cve, cyber, denial, deployment, dereference, disclosure, execute, exploit, hack, heap, identity, injection, interception, leak, overflow, privilege, remote, root, scripting, security, stack, threat, unauthenticated, vuln, xss \\
    \hline
    \end{tabular}
    \label{table::naive_words}
\end{table}

\paragraph{Classifier Configuration.}
Supervised machine learning techniques require design tailored to the problem at hand.
For each classifier employed, their relevant parameters and design variables were varied, namely the step size and the regularisation parameter (C) for the SVM, and the number of layers and neurons per layer for the MLP.
The size of the TF-IDF feature vector considered was also varied for both classifiers.
Through a Pareto-optimal search, ideal configurations were found: the best SVM uses a step size and C of 0.05 and 5, respectively, and the best MLP had 5 layers with 10 neurons each.
Both models use feature vectors with a size of 3000, revealing a clear advantage in using high-dimensional feature vectors.
A complete description of the methodology employed for the classifier's design can be found in Appendix~\ref{appendix-classifier-design}.

\paragraph{Clustering.}
\toolname{} uses the \kmeans{} algorithm in the offline clustering component, configured with fifty iterations, a minimum of two clusters, and the remaining parameters with their default values.
Clustering was performed on the set of tweets classified as positive.

The WTS cluster cohesion measure was set to $\tau=\frac{2}{3}$.
This value was selected after preliminary experiments, reflecting the rationale that two tweets can be in the same cluster if and only if they share at least two-thirds of their words.

We compare our data presentation strategy with the one employed by threat intelligence tools and SIEMs capable of collecting OSINT (\eg{}, AlienVault OTX~\cite{avotx}, Spiderfoot~\cite{spiderfoot}).
For that, we set up a Logstash~\cite{logstash} instance fed by the same dataset as \toolname{}, which selected as relevant tweets mentioning at least one of our infrastructure assets and containing at least one security concept.

The security concept keywords were selected using the following methodology.
First, a list of documents is obtained by selecting all tweets labelled as positive from all datasets.
After that, we removed stopwords, applied the TF-IDF method, and selected the words with TF-IDF value lower than a threshold $\rho$.
Finally, the list was manually filtered for security-irrelevant content (such as numbers).
We considered $\rho$ values of $0.1$, $0.2$, and $0.3$.
After inspecting the results, $\rho=0.2$ was chosen due to the provision of the most substantial amount of generic words without showing words related to a specific context.
The Logstash security concept keyword set corresponding to $\rho=0.2$ appears in Table~\ref{table::naive_words}.

For the time-window model we applied a $\theta$ value of seven days, \ie{}, a cluster without updates for seven days is removed from the online clustering state.
The same $\theta$ value was applied to the Logstash approach but globally, \ie{}, all relevant tweets were removed from the active threat pool after a week.

\section{Results}
\label{sec::results}

The tweet processing pipeline components were evaluated using the selected models and datasets \texttt{D2} and \texttt{D3}.
These consider only tweets in the future of those in the training set (\texttt{D1}), and include information posted by an additional and substantially larger set of accounts (\texttt{S2}) not considered in the training stage.
This methodology embodies the idea that in a real deployment, models will classify future tweets possibly from a different set of accounts.

Considering that 10-fold cross-validation was employed during the model selection phase, it should be noted that the selected model configurations were trained for the evaluation phase using the whole \texttt{D1} dataset.
The feature vectors of \texttt{D2} and \texttt{D3} tweets were generated using the TF-IDF model determined using dataset \texttt{D1}.
This guarantees that TF-IDF weights attributed to words in \texttt{D2} and \texttt{D3} will be coherent with those used to train the classifiers.

\subsection{Classification}
\label{sec::results-classification}

\begin{figure}[t]
    \centering
    \includegraphics[width=0.48\columnwidth]{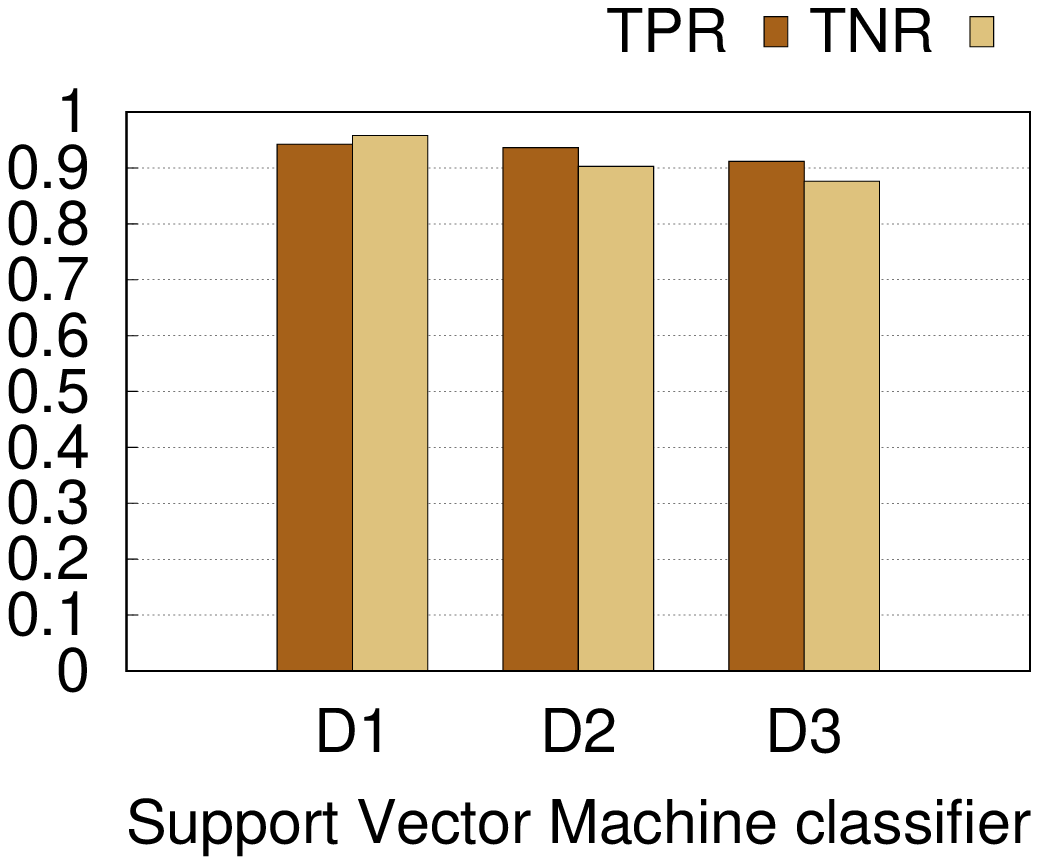}
    \includegraphics[width=0.48\columnwidth]{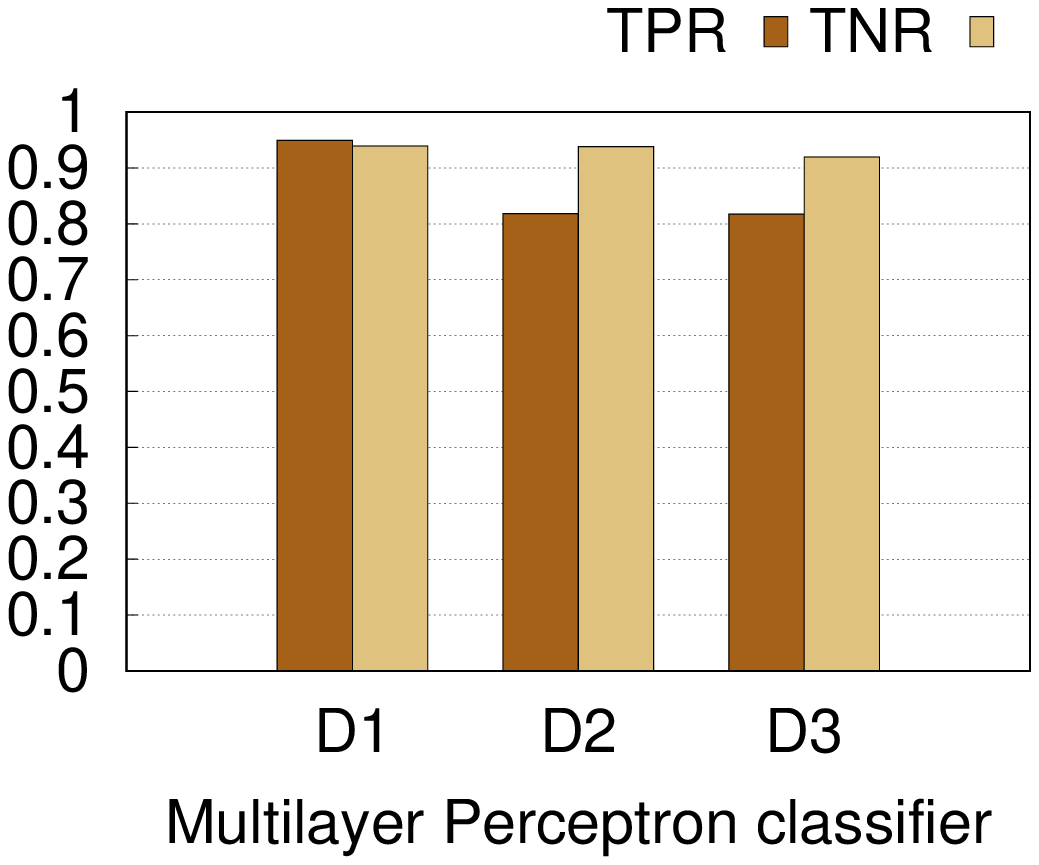}
    \caption{SVM (left) and MLP (right) classifier results.}
    \label{fig:classification_results}
\end{figure}

Figure~\ref{fig:classification_results} shows the True Positive Rate (TPR) and True Negative Rate (TNR) of the SVM and MLP classifiers described in Section~\ref{sec::setup}, considering also the average result of the 10-fold cross-validation over \texttt{D1}.

Overall, the results are slightly worse for \texttt{D2} and \texttt{D3} when compared to \texttt{D1} (as expected), since new data presents unmodeled patterns to the classifiers.
Focusing on the results obtained for \texttt{D2} and \texttt{D3}, in general, the classifiers maintain very high TPR and TNR, except for the MLP TPR.
In both cases, the TNR is higher than the TPR.
The imbalance between positively and negatively labelled data in the training data sets (more negative samples) can explain a higher TNR.

In summary, \emph{the SVM approach achieved the best results, displaying true positive and true negative rates around 90\% and showing a small degradation of results in \texttt{D2} and \texttt{D3}}.
For these reasons, the SVM model was employed in all further experiments.
These results support the application of a supervised classifier to select tweets relevant for cybersecurity.

\subsection{Clustering}
\label{sec::eval_clustering}

The following experiments evaluate \toolname{}'s ability to aggregate the dataflow into meaningful clusters, where each cluster is expected to describe a single threat.
Further, the DynamicClustream's window model is evaluated to assess its capability to detect the continuous discussion of threats.

The initial clustering evaluation focuses on the basic algorithm's capability of properly aggregate tweets, \ie{}, producing clusters with high internal cohesion and low inter-cluster similarity.
Then we analyse the end-to-end benefit of \toolname{} and discuss the effectiveness of the proposed outlier detection mechanism and time-window model which convey the active threat monitor functionality to \toolname.

Datasets \texttt{D2} and \texttt{D3} were merged and fed to \toolname{}.
At the end of each day, for all clusters in the current cluster state, we calculated the average WTS and the Jaccard distance between all pairs of clusters.
For the latter, we saved the largest value, which corresponds to the most similar cluster pair.
Since \toolname{}'s objective is to obtain distinct clusters, each devoted to a single threat, the WTS should always be high (\ie{}, the elements in each cluster are very similar), and the maximum Jaccard distance should be low (\ie{}, there are no clusters that should be merged).

\begin{figure}[t]
    \centering
    \includegraphics[width=\columnwidth]{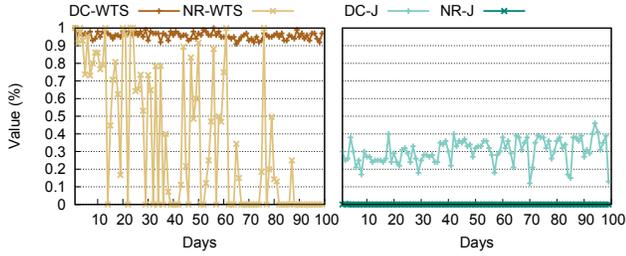}
    \caption{Comparing WTS and Jaccard distance over time, for DynamicClustream with and without the re-clustering step.}
    \label{fig:wts-jaccard}
\end{figure}

Figure~\ref{fig:wts-jaccard} shows the WTS and maximum Jaccard distance obtained, comparing the proposed DynamicClustream clustering algorithm (DC-WTS and DC-J) to its execution in clustering only mode, without considering re-clustering (NR-WTS and NR-J).
The importance of including the re-clustering step (lines 22-25 of Algorithm~\ref{alg::clustering}) is clear since it raises the WTS to above 90\% independently of the number of clusters and tweets present in the cluster state.
The Jaccard distance, although with small values, is higher when using the re-clustering algorithm.
Yet, this is an expected result.
First, re-clustering produces significantly more clusters, therefore naturally decreasing their degree of separation. 
Second, since tweets in clusters mentioning different threats are likely to share commonly used security concept words and sentence structure, their similarity is increased.

\begin{figure}[t]
    \centering
    \includegraphics[width=\columnwidth]{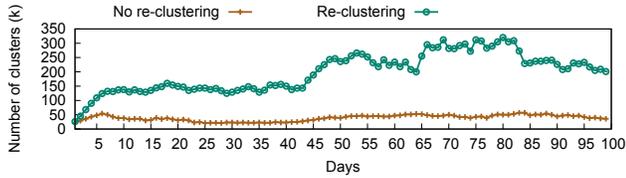}
    \caption{Number of clusters obtained by the DynamicClustream algorithm with and without the re-clustering step.}
    \label{fig:clustering-size}
\end{figure}

Regarding the number of clusters obtained using either approach, the re-clustering algorithm naturally increases the number of clusters, as shown in Figure~\ref{fig:clustering-size}.
Nevertheless, we argue that in practice, the DynamicClustream algorithm improves the balance between maximising the relevance of the information presented and minimising the time required for its analysis.
The WTS results provide guarantees that each cluster has similar tweets, likely about a single threat.
Therefore, we can be confident that the set of cluster exemplar tweets provides a complete and accurate summary of the current threat landscape, thus not requiring additional time to analyse more tweets.
Without the WTS cohesion validation, each cluster may discuss various threats---a highly plausible assumption based on the very low WTS values in Figure~\ref{fig:wts-jaccard} for the NR-WTS case---meaning that all tweets of each cluster would have to be analysed.


\begin{figure}[t]
    \centering
    \includegraphics[width=\columnwidth]{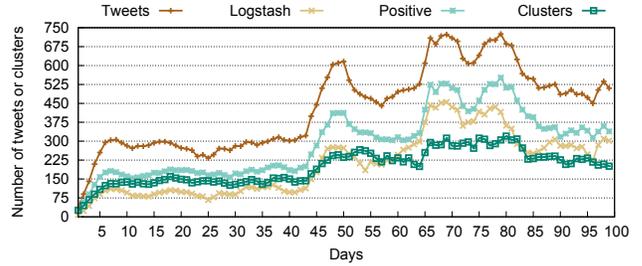}
    \caption{The number of tweets collected and those filtered by Logstash, classification only, and classification and clustering.}
    \label{fig:presentation_comparison}
\end{figure}
\begin{figure}[t]
    \centering
    \includegraphics[width=\columnwidth]{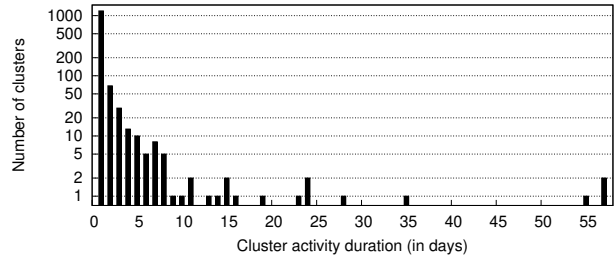}
    \caption{The distribution of the number of clusters over the cluster duration in days.}
    \label{fig:cluster_sizes_days}
\end{figure}

\paragraph{End-to-end Benefit.}
The results presented in Figure~\ref{fig:presentation_comparison} highlight the end-to-end benefit of using \toolname{}, and reinforces the importance of its clustering stage.
The figure shows the reduction in the number of tweets that have to be analysed, when compared to the tweet stream, to the classifier output and to the naive Logstash filter described in Section~\ref{sec::setup}.

The results show the need for efficient OSINT retrieval tools.
Even with the naive keyword-based approach provided by the Logstash filter, the number of tweets marked as relevant would be extremely high, rendering the approach useless to SOC analysts.
The introduction of a trained classifier decreases the amount of information by 65\%.
By attaching a clustering stage, we further reduce the information to be shown by almost 80\%, which is a significant improvement.

\begin{table*}[!ht]
\caption{Examples of tweets whose content has high impact or important actionability.}
\label{tab::actionability-example}
\scriptsize
\setlength\tabcolsep{3pt}
\begin{tabularx}{\textwidth}{|p{0.35\textwidth}|P{0.015\linewidth}|P{0.04\textwidth}|P{0.03\textwidth}|P{0.04\textwidth}|P{0.08\textwidth}|X|}
\hline
\textbf{Cluster exemplar     text (without links)}&    \textbf{\#}    &    \textbf{Asset}    &    \textbf{Date}    &    \textbf{Action}    &    \textbf{Threat type}    &    \textbf{Notes}    \\\hline
\#ubuntu \#security : USN-3006-1: Linux kernel vulnerabilities                                                                    &        19        &        Linux        &        10/06            &        Patch        &    vulnerabilities                                            &    Several vulnerabilities patched, some were not yet included on NVD, half with CVSS $\geq 7.0$    \\\hline
High - USN-3016-1 - Linux kernel vulnerabilities A security issue affects these releases of Ubuntu and its derivat                    &        12        &        Linux        &        27/06            &        Patch        &    vulnerabilities                                            &    Several vulnerabilities patched, some were not yet included on NVD, half with CVSS $\geq 7.5$    \\\hline
Microsoft Internet Explorer CVE-2016-3205 Scripting Engine Remote Memory Corruption Vulnerability Type: Vulnerabil                    &        8        &        IE            &        14/06 (1)        &        Config        &    vulnerability, remote                                    &    This cluster contains various threats with CVSS $\geq 7.5$; configurations are suggested to mend the issue before it is patched    \\\hline
\#CISCO fixed severe \#vulnerabilities in Network Management and \#Security Products \#SecurityAffairs                                &        9        &        Cisco        &        30/06 (2)        &        Patch        &    vulnerabilities                                            &    Patch for critical vulnerabilities (CVSS $\geq 8.6$) announced on Twitter before being published on NVD        \\\hline
Bugtraq: {[}security bulletin{]} - Linux Kernel Flaw, ASN.1 DER decoder for x509 certificate DER                                    &        6        &        Linux        &        06/06 (21)        &        Patch        &    certificate                                &        A highly important Linux kernel flaw (CVSS $7.8$) was disclosed 21 days before being included in NVD    \\\hline
Vuln: Oracle Java SE and JRockit CVE-2016-3427 Remote Security Vulnerability Vulnerable:Red Hat Enterprise Linux                    &        21        &        Oracle        &        05/07            &        Patch        &    vulnerability, remote                                    &    This cluster contains three different threats (one with CVSS $9.0$); patches are available    \\\hline
Bugtraq: Cisco Security Advisory: Cisco RV110W, RV130W, and RV215W Routers Arbitrary Code Execution Vulnerability                    &        5        &        Cisco        &        15/06 (3)        &        Patch        &    vulnerability, execution                    &        A critical vulnerability (CVSS $9.8$) was disclosed and patched before its inclusion on NVD    \\\hline
Bugtraq: Cisco Security Advisory: Cisco Products IPv6 Neighbor Discovery Crafted Packet Denial of Service                             &        5        &        Cisco        &        25/05 (4)        &        Patch        &    denial of service                        &        A high impact vulnerability (CVSS $7.5$)    was disclosed and patched before its inclusion on NVD    \\\hline
\#ubuntu \#security : USN-2975-2: Linux kernel (Trusty HWE) vulnerability                                                            &        5        &        Linux        &        16/05 (42)        &        Patch        &    vulnerability                            &        A high impact vulnerability (CVSS $7.8$)    was disclosed and patched before its inclusion on NVD (42 days in advance)    \\\hline
Bugtraq: Wordpress Levo-Slideshow 2.3 - Arbitrary File Upload Vulnerability                                                        &        9        &        WPress        &        07/06            &        Config        &    vulnerability                                            &    An exploit is provided; a software correction is suggested    \\\hline
\end{tabularx}
\end{table*}

\paragraph{Active Threat Monitor.}
To demonstrate the necessity of the active threat monitor implemented by the proposed stream clustering algorithm, we measured the active time for each of the 820 clusters formed during \toolname{}'s operation on the union of datasets \texttt{D2} and \texttt{D3}.
We define the duration of a cluster as the difference in days between the date of its creation and the date of the last added tweet.
Figure~\ref{fig:cluster_sizes_days} depicts the distribution of the number of clusters over the cluster duration in days.
The results clearly show that a global time-window model enforcing a fixed duration for each tweet would fail to detect active topics through time, since the threat discussion duration varies greatly (between 1 and 57 days), even in a dataset that covers only 100 days.

\subsection{Analysis of Generated IoCs}
\label{sec::IOC_analysis}

Besides the ability to accurately select and aggregate tweets relevant to the security of an IT infrastructure, \toolname{} provides useful threat intelligence for SOC analysts.
To demonstrate this, we present some information about the timeliness, actionability, and relevance of the IoCs generated from the dataset used in previous experiments.

From the data collected over 3 months, \toolname{} generated 820 clusters (IoCs) containing 1754 tweets.
From these, we selected those with 5 or more tweets for analysis, obtaining 65 clusters comprising 466 tweets.
These clusters are listed in Appendix~\ref{appendix-tweet-table}.
The remaining 755 clusters have 1 (577 clusters), 2 (101), 3 (55), and 4 tweets (22).
Our focus on larger clusters was motivated by the expectation that relevant threats are probably those that attract more attention and, ultimately, are mentioned in more tweets.

All tweets within each cluster were manually analysed.
From these, as well as from any hyperlink therein, we extracted all CVEs mentioned (if any) and their Common Vulnerability Scoring System v3.0 (CVSS)~\cite{cvss} impact score, the types of actions that can be performed to respond to the alarm, and a comparison between the date of the earliest tweet in the cluster and the CVE's publication date on NVD.

The actionability information was divided into three categories: a patch is available (45 occurrences); a configuration to avoid the vulnerability exploitation is suggested (2 occurrences); and no directly actionable information is provided (14 occurrences).
The latter is mostly associated with clusters mentioning exploits to vulnerabilities, with the tweet hyperlinks leading to proofs-of-concept.
However, an expert might still make use of this information to prevent exploitation, as discussed in previous work~\cite{sabottke2015vulnerability}.
Patches are mostly announced together with their associated vulnerabilities, regardless of indexing on NVD.
In the end, 71\% (46) of the clusters provided directly usable intelligence, including exploits whose vulnerabilities were not matched to NVD entries.

Among the 65 clusters, 36 mentioned a total of 122 different CVEs (15 clusters mentioning more than one CVE).
Of these, only two have low impact score, about a quarter have medium impact (33), more than half are categorised with high impact (68), and more than a tenth have critical impact (14).

Considering their relevance, 43\% (28) of the IoCs were related to CVSS scores above or equal to 7 (high severity) and 12\% (8) to scores above or equal to 9 (critical severity).
Regarding timeliness, 20\% of the alerts were raised 8 days (on average) before their corresponding vulnerabilities were published on NVD.

As an illustration of the richness of the obtained data, Table~\ref{tab::actionability-example} shows 10 representative IoCs selected from those analysed.
In the table, the date column shows the date of the earliest tweet in the cluster and, when a number is shown within parenthesis, it denotes the number of days before publication on NVD.
Two additional columns provide information about the threat type (as automatically classified by \toolname{}) and relevant notes about the cluster content.

From the 10 clusters presented, 6 announce vulnerabilities before publication on NVD, all of them with patches available.
Further, 7 are classified with a \emph{high} CVSS and two with \emph{critical} impact.
For example, the \nth{7} IoC of the table shows a critical Cisco router vulnerability patched and published three days before its inclusion on NVD.
Finally, since not all occurrences are patched at disclosure time, some actionable IoCs contain suggested configurations to avoid exploitations. As an example, the last row in the table shows a WordPress exploit with suggested remediations.

These results show the edge obtained by using Twitter as a security data source.
A SOC analyst using \toolname{} would obtain timely and relevant data about patches to known vulnerabilities, thus possibly reducing the vulnerable system's exposure time.
Further, the results also show that vendors publish important impact data before it is included in NVD.

\section{Conclusions}
\label{sec::conclusions}

This paper proposes \toolname{}, a Twitter-based streaming threat monitor for threat detection in security operation centres.
It implements a pipeline that gathers tweets from a set of accounts, filters them based on the monitored infrastructure, and classify the remaining tweets as either relevant or not.
Relevant tweets are grouped in dynamic clusters and presented as indicators of compromise that can be either manually inspected or fed to SIEMs and other threat intelligence tools.
Results show that our system maximises the relevant information (true positive rate of 90\%), minimises irrelevant information (false positive rate of 10\%), and aggregates related information (only 21\% of the relevant tweets are presented).
Finally, we performed an evaluation of the IoCs generated by \toolname, showing that highly relevant, timely and actionable information was collected, illustrating the value of our end-to-end approach.

\paragraph{Acknowlegments.}
We thank Andr\'e Correia for collecting and labelling the data set employed in this paper.  
This work was partially supported by the EC through funding of the H2020 DiSIEM project (H2020-700692), and by the LASIGE Research Unit (UID/CEC/00408/2019).

\bibliographystyle{abbrv}
\bibliography{main}

\newpage

\appendix
\FloatBarrier

\section{Pareto figures}
\label{appendix-classifier-design}

\begin{figure}[t]
    \centering
    \includegraphics[width=\linewidth]{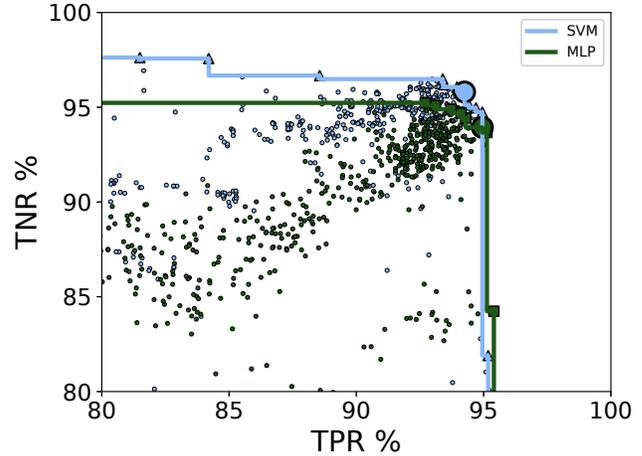}
    \caption{The Pareto fronts for SVM and MLP cross-validated using \textit{D1}.}
    \label{fig:pareto}
\end{figure}

\paragraph{Feature extraction.}

We used Spark's implementation of TF-IDF with default parameters, except for the feature vector size.
In order to find a suitable vector size to describe the tweets, eleven values were tested: $\{30, 50, 80,$ $100, 200, 300, 500, 750, 1000, 1500, 3000\}$.
This range covers from low to high dimensional vectors, and with it, we should be able to find an appropriate vector size for the datasets.

\paragraph{Classification.}

As mentioned in Section~\ref{sec::methodology}, two classifiers were employed: a linear SVM and an MLP Neural Network.
Relevant hyper-parameters and design variables were varied to find a good design for this application.
For the SVM, we varied $C$ (the regularization parameter) within $\{0.01, 0.02, 0.05,$ $0.1, 0.2, 0.5, 1, 2, 5\}$, and the step size (a parameter for the Stochastic Gradient Descent)
within $\{0.1, 0.5, 1, 1.5, 2, 5\}$.
For the MLP, the number of layers varied from $2$ to $8$ and the number of neurons per layer within $\{5, 7, 10, 12, 14, 16, 18, 20\}$.

Each model was evaluated through a 10-fold cross-validation procedure using dataset \texttt{D1}.
The maximum number of training iterations was set to 100 for the SVM and 200 for the MLP, which were deemed to achieve parameter convergence for the range of the design parameters.

To select the best classifiers, we performed a Pareto-optimal search.
For each type of classifier we plotted a Pareto front figure (Figure~\ref{fig:pareto}), with lines connecting the dominant configurations regarding \emph{True Positive Rate} (TPR, x-axis) and \emph{True Negative Rate} (TNR, y-axis).
Each point shows the average value obtained by a specific configuration over the 10-fold cross-validation procedure.
The highlighted triangular and circular points are, respectively, the dominant configurations and the configurations chosen to be used (the SVM case) in the experiments. We use the classical true positive definition: a sample labelled as positive and classified as positive;
in our case, a tweet manually labelled as relevant and classified as relevant.
The negative samples use the equivalent definition.

Based on this analysis we select the configurations with the best TPR$\times$TNR balance: those with the smallest distance to the optimum.
The best SVM configuration uses a step size and C values of 0.05 and 5, respectively, and the best MLP had 5 layers with 10 neurons each.
Both models use feature vectors with a size of 3000, revealing a clear advantage in using high-dimensional feature vectors.

\section{Complete Cluster Data}
\label{appendix-tweet-table}

Tables~\ref{tab::actionability1} and~\ref{tab::actionability2} present the 65 IoCs largest clusters generated by \toolname{}, as described in Section~\ref{sec::eval_clustering}.

By running \toolname{}'s IoC generation module, each cluster was tagged with the type of threats mentioned by its tweets.
The most common tags are ``vulnerability'' (23) and ``vulnerabilities'' (13), reflecting that most threats are related to vulnerability disclosure.
Other two common tags are ``exploit'' (18) and ``0day'' (15) (or ``zero-day''), which indicate exploitable vulnerabilities.
Less used tags include ``remote'' (6) (remote execution attacks), ``denial of service'' (6), ``SQL injection'' (5), and ``Buffer overflow'' (4) (or BO).

\begin{figure}[t]
    \centering
    \includegraphics[width=\columnwidth]{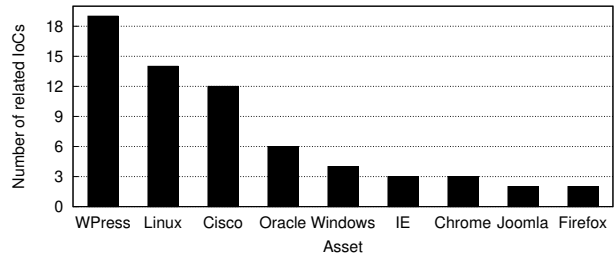}
    \caption{Number of IoCs for each asset.}
    \label{fig:assets}
\end{figure}

Out of the 13 assets composing the hypothetical IT infrastructure described in Table~\ref{table::infrastructures}, only 9 ($\sim70\%$) had related IoCs.
The distribution of IoCs over the assets is shown in Figure~\ref{fig:assets}.
WordPress is the asset with more related IoCs (19), followed by Linux (14) and Cisco (12).
All analysed IoCs mentioned a single asset.

\begin{table*}
\caption{Largest generated clusters represented as IoCs.}
\label{tab::actionability1}
\scriptsize
\setlength\tabcolsep{3pt}
\begin{tabularx}{\textwidth}{|p{0.35\textwidth}|P{0.01\linewidth}|P{0.05\linewidth}|P{0.03\linewidth}|P{0.035\linewidth}|P{0.17\textwidth}|p{0.265\textwidth}|}

\hline
\textbf{Cluster exemplar	 text (without links)}&	\textbf{\#}	&	\textbf{Asset}	&	\textbf{Date}	&	\textbf{Action}	&	\textbf{Threat type}	&	\textbf{Notes}	\\\hline
Vuln: Oracle Java SE and JRockit CVE-2016-3427 Remote Security Vulnerability Vulnerable:Red Hat Enterprise Linux					&		21		&		Oracle		&		05/07			&		Patch		&	vulnerability, remote									&	This cluster contains three different threats (one with CVSS $9.0$); patches are available	\\\hline
\#ubuntu \#security : USN-3006-1: Linux kernel vulnerabilities																	&		19		&		Linux		&		10/06			&		Patch		&	vulnerabilities											&	Several vulnerabilities patched, some were not yet included on NVD, half with CVSS $\geq 7.0$	\\\hline
\#0daytoday \#Cisco EPC 3928 - Multiple Vulnerabilities  {[}webapps \#exploits \#Vulnerabilities \#0day \#Exploit{]}				&		16		&		Cisco		&		07/06			&		None			&	exploit, vulnerabilities, 0day							&	An exploit is presented; an expert might use this data for protection (half of the vulns with CVSS $\geq 7.5$)	\\\hline
\#0daytoday \#Joomla En Masse com\_enmasse Component 5.1 - 6.4 - SQL Injection Vulnerability {[}\#0day \#Exploit{]}				&		12		&		Joomla		&		15/06			&		None			&	SQL injection, exploit, injection, vulnerability, 0day	&	An exploit is presented; an expert might use this data for protection	\\\hline
High - USN-3016-1 - Linux kernel vulnerabilities A security issue affects these releases of Ubuntu and its derivat					&		12		&		Linux		&		27/06			&		Patch		&	vulnerabilities											&	Several vulnerabilities patched, some were not yet included on NVD, half with CVSS $\geq 7.5$	\\\hline
\#0daytoday \#Sun Secure Global Desktop and Oracle Global Desktop 4.61.915 - ShellShock Exploit {[}\#0day \#Exploit{]}				&		11		&		Oracle		&		06/06			&		None			&	exploit, 0day											&	An exploit is presented; an expert might use this data for protection	\\\hline
\#ubuntu \#security : USN-2993-1: Firefox vulnerabilities																			&		10		&		Firefox		&		09/06 (4)		&		Patch		&	vulnerabilities											&	Patches are available for vulnerabilities, half with CVSS $\geq 8.8$	\\\hline
Bugtraq: CM Ad Changer 1.7.7 Wordpress Plugin - Cross Site Scripting Web Vulnerability											&		10		&		WPress		&		13/06			&		Patch		&	vulnerability											&	A patch is available; an exploit is provided	\\\hline
Bugtraq: Wordpress Levo-Slideshow 2.3 - Arbitrary File Upload Vulnerability														&		9		&		WPress		&		07/06			&		Config		&	vulnerability											&	An exploit is provided; a software correction is suggested	\\\hline
Bugtraq: Oracle Orakill.exe Buffer Overflow																						&		9		&		Oracle		&		14/06 			&		Patch		&	Buffer overflow											&	A patch is available; an exploit is provided	\\\hline
\#CISCO fixed severe \#vulnerabilities in Network Management and \#Security Products \#SecurityAffairs								&		9		&		Cisco		&		30/06 (2)		&		Patch		&	vulnerabilities											&	Patch for critical vulnerabilities (CVSS $\geq 8.6$) announced on Twitter before being published on NVD		\\\hline
\#ubuntu \#security : USN-3016-1: Linux kernel vulnerabilities																	&		8		&		Linux		&		27/06			&		Patch		&	vulnerabilities											&	Several vulnerabilities patched, some were not yet included on NVD, half of the vulns with CVSS $\geq 7.5$		\\\hline
Microsoft Internet Explorer CVE-2016-3205 Scripting Engine Remote Memory Corruption Vulnerability Type: Vulnerabil					&		8		&		IE			&		14/06 (1)		&		Config		&	vulnerability, remote									&	This cluster contains various threats with CVSS $\geq 7.5$; configurations are suggested to solve the issue before it is patched	\\\hline
NA - CVE-2016-2825 - Mozilla Firefox before 47.0 allows remote... Mozilla Firefox before 47.0 allows remote attack					&		8		&		Firefox		&		13/06			&		Patch		&	attack, remote											&	A patch is available for a vulnerability with CVSS $6.5$		\\\hline
\#0daytoday \#WordPress Social Stream Plugin 1.5.15 - wp\_options Overwrite Vulnerability {[}\#0day \#Exploit{]}					&		8		&		WPress		&		14/06			&		Patch		&	exploit, vulnerability, 0day								&	A patch is available; an exploit is provided	\\\hline
Microsoft Internet Explorer 11 Garbage Collector Attribute Type Confusion \#exploit												&		8		&		IE			&		18/06			&		Patch		&	exploit													&	A patch is available for a vulnerability with CVSS $8.8$; an exploit is provided	\\\hline
CVE-2016-1388 Cisco Prime Network Analysis Module (NAM) before 6.1(1) patch.6.1-2-final and 6.2.x before 6.2(1) an					&		8		&		Cisco		&		03/06			&		Patch		&															&	This cluster contains 4 threats, 3 with CVSS $\geq 7.8$; patches are available	\\\hline
\#Oracle \#Linux 6 : \#openssl (ELSA-2016-0996) \#Nessus																			&		8		&		Linux		&		16/05			&		Patch		&															&	This cluster contains seven threats: 3 critical (CVSS $9.8$) and 3 high (CVSS $7.5$); patches are available	\\\hline
Vuln: Linux Kernel Multiple Local Memory Corruption Vulnerabilities 																&		7		&		Linux		&		08/07 			&		Patch		&	vulnerabilities											&	Patches are available for vulnerabilities with CVSS $7.1$ and $7.8$	\\\hline
Vuln: Linux Kernel CVE-2016-0723 Local Race Condition Vulnerability 																&		7		&		Linux		&		08/07			&		Patch		&	vulnerability											&	A patch is available	for vulnerability with CVSS $6.8$	\\\hline
Vuln: Linux kernel CVE-2013-7446 Use After Free Denial of Service Vulnerability 													&		7		&		Linux		&		05/07			&		Patch		&	denial of service, vulnerability							&	A patch is available	for vulnerability with CVSS $5.3$	\\\hline
Bugtraq: Cisco Security Advisory: Cisco Firepower System Software Static Credential Vulnerability					 				&		7		&		Cisco		&		29/06 (3)		&		Patch		&	vulnerability											&	A patch is available	for vulnerability with CVSS $8.6$	\\\hline
\#0daytoday \#WordPress Ultimate Membership Pro Plugin 3.3 - SQL Injection Vulnerability {[}\#0day \#Exploit{]}					&		7		&		WPress		&		29/06			&		Patch		&	SQL injection, exploit, injection, vulnerability, 0day	&	A patch is available; an exploit is provided	\\\hline
\#0daytoday \#Google Chrome - GPU Process MailboxManagerImpl Double-Read Vulnerability {[}\#0day \#Exploit{]}						&		7		&		Chrome		&		15/06			&		Patch		&	exploit, vulnerability, 0day								&	A patch is available; an exploit is provided	\\\hline
\#0daytoday \#WordPress Gravity Forms Plugin 1.8.19 - Arbitrary File Upload Exploit {[}\#0day \#Exploit{]}							&		7		&		WPress		&		17/06			&		None			&	exploit, 0day											&	An exploit is presented; an expert might use this data for protection	\\\hline
\#0daytoday \#WordPress Uncode Theme 1.3.1 - Arbitrary File Upload Exploit {[}webapps \#exploits  \#0day \#Exploit{]}				&		7		&		WPress		&		06/06			&		N/A			&	exploit, 0day											&	All tweet links are broken; nothing can be inferred	\\\hline
\#0daytoday \#WordPress Double Opt-In for Download Plugin 2.0.9 - SQL Injection Vulnerability {[}\#0day \#Exploit{]}				&		7		&		WPress		&		06/06			&		Patch		&	SQL injection, exploit, injection, vulnerability, 0day	&	A patch is available; an exploit is provided	\\\hline
\#cybersecurity Hackers offering Microsoft Windows zero-day exploit for \$90000 \#infosec											&		7		&		Windows		&		01/06 			&		N/A			&	exploit													&	Just informative tweets	\\\hline
\#Oracle ATS Arbitrary File Upload \#PacketStorm																					&		7		&		Oracle		&		24/05			&		None			&															&	An exploit is presented; an expert might use this data for protection	\\\hline
Vuln: Linux Kernel 'usb/core/hub.c' NULL Pointer Dereference Denial of Service Vulnerability										&		6		&		Linux		&		08/07			&		Patch		&	denial of service, vulnerability							&	A patch is available	for vulnerability with CVSS $6.8$	\\\hline
\#0daytoday \#Linux - ecryptfs and /proc/\$pid/environ Privilege Escalation Vulnerability {[}\#0day \#Exploit{]}					&		6		&		Linux		&		21/06 (6)		&		None			&	exploit, escalation, vulnerability, 0day					&	An exploit is early presented for a vulnerability with CVSS $7.8$; an expert might use this data for protection	\\\hline
CVE-2016-3221 The kernel-mode drivers in Microsoft Windows Vista SP2, Windows Server 2008 SP2 and R2 SP1, Windows					&		6		&		Windows		&		16/06			&		Patch		&															&	A patch is available	for a vulnerability with CVSS $7.8$\\\hline
NA - CVE-2016-3201 - Microsoft Windows 8.1, Windows Server 2012 Gold... Microsoft Windows 8.1, Windows Server 2012					&		6		&		Windows		&		16/06 			&		Patch		&															&	A patch is available	for a vulnerability with CVSS $6.5$\\\hline
\end{tabularx}
\end{table*}

\begin{table*}
\caption{Largest generated clusters represented as IoCs (cont.).}
\label{tab::actionability2}
\scriptsize
\setlength\tabcolsep{3pt}
\begin{tabularx}{\textwidth}{|p{0.35\textwidth}|P{0.01\linewidth}|P{0.05\linewidth}|P{0.03\linewidth}|P{0.035\linewidth}|P{0.17\textwidth}|p{0.265\textwidth}|}

\hline
\textbf{Cluster exemplar	 text (without links)}&	\textbf{\#}	&	\textbf{Asset}	&	\textbf{Date}	&	\textbf{Action}	&	\textbf{Threat type}	&	\textbf{Notes}	\\\hline
\#0daytoday \#Joomla com\_affiliatetracker - SQL Injection Vulnerability {[}webapps \#exploits \#Vulnerability \#0day				&		6		&		Joomla		&		13/06			&		N/A			&	SQL injection, exploit, injection, vulnerability, 0day	&	All tweet links are broken; nothing can be inferred	\\\hline
{[}shellcode{]} - \#Linux x86\_64 Shellcode Null-Free Reverse TCP Shell \#ExploitDB												&		6		&		Linux		&		16/06 			&		None			&	exploit													&	An exploit is presented; an expert might use this data for protection	\\\hline
Bugtraq: {[}security bulletin{]} - Linux Kernel Flaw, ASN.1 DER decoder for x509 certificate DER									&		6		&		Linux		&		06/06 (21)		&		Patch		&	certificate								&		A highly important Linux kernel flaw (CVSS $7.8$) was disclosed 21 days before being included in NVD	\\\hline
{[}webapps{]} - WordPress WP Mobile Detector Plugin 3.5 - Arbitrary File Upload: WordPress WP Mobile Detector Plu...				&		6		&		WPress		&		06/06			&		Patch		&											&		A patch is available; an exploit is provided	\\\hline
Bugtraq: Cisco Security Advisory: Cisco Prime Network Analysis Module IPv6 Denial of Service Vulnerability							&		6		&		Cisco		&		01/06 (1)		&		Patch		&	denial of service, vulnerability			&		A patch is available	for a vulnerability with CVSS $5.3$	\\\hline
Bugtraq: Cisco Security Advisory: Cisco Prime Network Analysis Module Unauthenticated Remote Code \#bugtraq						&		6		&		Cisco		&		01/06 (1)		&		Patch		&	remote									&		A patch is available	for a critical vulnerability with CVSS $9.8$	\\\hline
WordPress Patches Zero Day in WP Mobile Detector Plugin \#InfoSec																	&		6		&		WPress		&		03/06 			&		Patch		&	zero day									&		A patch is available		\\\hline
CVE-2016-1381 Memory leak in Cisco AsyncOS 8.5 through 9.0 before 9.0.1-162 on Web Security Appliance (WSA) device					&		6		&		Cisco		&		25/05			&		Patch		&	leak										&		A patch is available	for a vulnerability with CVSS $7.5$	\\\hline
Oracle E-Business Suite Vulnerabilities Related To Common Components Oracle E-Business Intelligence component in O					&		6		&		Oracle		&		23/05			&		None			&	vulnerabilities							&		The tweet links provide no useful information	\\\hline
NA - cisco-sa-20160518-wsa4 - Cisco Web Security Appliance Connection Denial of Service Vulnerability A vulnerabil					&		6		&		Cisco		&		18/05 (6)		&		Patch		&	denial of service, vulnerability			&		A high impact vulnerability (CVSS $7.5$)	was disclosed and patched before its inclusion on NVD	\\\hline
\#ubuntu \#security : USN-2947-1: Linux kernel vulnerabilities																	&		6		&		Linux		&		06/04			&		Patch		&	vulnerabilities							&		A patch is available	to solve multiple vulnerabilities, one of them critical (CVSS $9.8$)	\\\hline
Vuln: Cisco Video Communication Server and Expressway CVE-2016-1444 Authentication Bypass Vulnerability							&		5		&		Cisco		&		08/07			&		Patch		&	vulnerability							&		A patch is available	for a vulnerability with CVSS $6.5$	\\\hline
Vuln: Google Chrome Prior to 49.0.2623.75 Multiple Security Vulnerabilities														&		5		&		Chrome		&		06/07			&		Patch		&	vulnerabilities							&		A patch is available to solve multiple high to critical vulnerabilities (5 with CVSS $8.8$ and 5 with CVSS $9.8$) \\\hline 
{[}webapps{]} - WordPress Real3D FlipBook Plugin - Multiple Vulnerabilities: WordPress Real3D FlipBook Plugin - M...				&		5		&		WPress		&		04/07			&		None			&	vulnerabilities							&		An exploit is presented; an expert might use this data for protection	\\\hline
Vuln: Linux Kernel 'btrfs/inode.c' Information Disclosure Vulnerability															&		5		&		Linux		&		05/07			&		Patch		&	vulnerability							&		A patch is available for a vulnerability with CVSS $4.0$ \\\hline 
Medium - CVE-2016-5835 - WordPress before 4.5.3 allows remote attackers... WordPress before 4.5.3 allows remote at					&		5		&		WPress		&		29/06			&		Patch		&	attack, remote							&		A patch is available	for a vulnerability with CVSS $7.5$	\\\hline
\#vulnerability \#security : WordPress Contus Video Comments 1.0 File Upload														&		5		&		WPress		&		22/06			&		None			&	vulnerability							&		An exploit is presented; an expert might use this data for protection	\\\hline
{[}webapps{]} - WordPress Ultimate Product Catalog Plugin 3.8.1 - Privilege Escalation: WordPress Ultimate Produc...				&		5		&		WPress		&		20/06 			&		Patch		&	escalation								&		A patch is available; an exploit is provided	\\\hline
\#0daytoday \#WordPress Premium SEO Pack 1.9.1.3 - wp\_options Overwrite Exploit {[}webapps \#exploits  \#0day \#Exploit{]}		&		5		&		WPress		&		21/06			&		None			&	exploit, 0day							&		An exploit is presented; an expert might use this data for protection	\\\hline
CVE-2016-0200 Microsoft Internet Explorer 9 through 11 allows remote attackers to execute arbitrary code or cause			 		&		5		&		IE			&		16/06			&		Patch		&	attack, remote							&		The cluster contains two different threats; patches are available to solve 4 vulns with CVSS $8.8$	\\\hline
Bugtraq: Cisco Security Advisory: Cisco RV110W, RV130W, and RV215W Routers Arbitrary Code Execution Vulnerability					&		5		&		Cisco		&		15/06 (3)		&		Patch		&	vulnerability, execution					&		A critical vulnerability (CVSS $9.8$) was disclosed and patched before its inclusion on NVD	\\\hline
\#0daytoday \#WordPress Newspaper Theme 6.7.1 - Privilege Escalation Exploit {[}webapps \#exploits  \#0day \#Exploit{]}			&		5		&		WPress		&		06/06			&		Patch		&	exploit, escalation, 0day				&		A patch is available; an exploit is provided	\\\hline
{[}webapps{]} - WordPress Simple Backup Plugin 2.7.11 - Multiple Vulnerabilities: WordPress Simple Backup Plugin ...				&		5		&		WPress		&		06/06			&		None			&	vulnerabilities							&		An exploit is presented; an expert might use this data for protection	\\\hline
CVE-2016-1701 The Autofill implementation in Google Chrome before 51.0.2704.79 mishandles the interaction between 					&		5		&		Chrome		&		06/06			&		Patch		&											&		All tweets refer a different vulnerability, all from the same date, all with CVSS $\geq 7.5$; patches are available	\\\hline
\#0daytoday \#WordPress WP PRO Advertising System Plugin 4.6.18 - SQL Injection Exploit {[}\#0day \#Exploit{]}						&		5		&		WPress		&		06/06			&		None			&	SQL injection, exploit, injection, 0day	&		An exploit is presented; an expert might use this data for protection	\\\hline
{[}webapps{]} - WordPress Creative Multi-Purpose Theme 9.1.3 - Stored XSS: WordPress Creative Multi-Purpose Theme...				&		5		&		WPress		&		06/06			&		Patch		&	XSS										&		A patch is available; an exploit is provided	\\\hline
\#WordPress WP Mobile Detector 3.5 Shell Upload \#PacketStorm																		&		5		&		WPress		&		04/06 			&		Patch		&											&		A patch is available; an exploit is provided	\\\hline
\#hackers Selling Unpatched Microsoft Windows Zero-Day Exploit for \$90.000														&		5		&		Windows		&		03/06 			&		N/A			&	exploit									&		Just informative tweets	\\\hline
Oracle E-Business Suite  Vulnerabilities Related To  E-Business Intelligence Oracle E-Business Intelligence compon					&		5		&		Oracle		&		30/05 			&		None			&	vulnerabilities							&		The tweet links provide no useful information	\\\hline
Bugtraq: Cisco Security Advisory: Cisco Products IPv6 Neighbor Discovery Crafted Packet Denial of Service 							&		5		&		Cisco		&		25/05 (4)		&		Patch		&	denial of service						&		A high impact vulnerability (CVSS $7.5$)	was disclosed and patched before its inclusion on NVD	\\\hline
\#ubuntu \#security : USN-2975-2: Linux kernel (Trusty HWE) vulnerability															&		5		&		Linux		&		16/05 (42)		&		Patch		&	vulnerability							&		A high impact vulnerability (CVSS $7.8$)	was disclosed and patched before its inclusion on NVD (42 days in advance)	\\\hline
Bugtraq: Cisco Security Advisory: Cisco Web Security Appliance HTTP POST Denial of Service Vulnerability							&		5		&		Cisco		&		18/05 (6)		&		Patch		&	vulnerability							&		A high impact vulnerability (CVSS $7.5$)	was disclosed and patched before its inclusion on NVD	\\\hline

\end{tabularx}
\end{table*}

\end{document}